\documentclass[aps,prd,twocolumn,nofootinbib,10pt]{revtex4-2}

\usepackage{graphicx}
\usepackage{color}
\usepackage{amsmath,amssymb}
\usepackage{mathtools}
\usepackage{physics}
\usepackage{bbold}
\usepackage{t1enc}
\usepackage{yhmath}
\usepackage{amsfonts}
\usepackage{latexsym}
\usepackage{amsfonts}
\usepackage{dsfont}
\usepackage{bbold}
\usepackage{latexsym} 
\usepackage{cancel}
\usepackage{graphicx, rotating}
\usepackage{epsfig}
\usepackage{color}
\usepackage[dvipsnames]{xcolor}
\usepackage{multirow}
\usepackage{color}
\usepackage[normalem]{ulem}
\usepackage{accents}
\usepackage{float}

\newcommand{\bea}{\begin{eqnarray}}
\newcommand{\eea}{\end{eqnarray}}
\newcommand{\be}{\begin{equation}}
\newcommand{\ee}{\end{equation}}
\newcommand{\pbar}[1]{\accentset{(-)}{#1}}

\DeclareMathSymbol{\shortminus}{\mathbin}{AMSa}{"39}
\DeclareMathSymbol{\shm}{\mathbin}{AMSa}{"39}

\def\simlt{\stackrel{<}{{}_\sim}}

\begin{document}

\title{Understanding Bell locality tests at colliders }

\begin{center}
\begin{flushright}
IFT-UAM/CSIC-26-33
\end{flushright}
\end{center}
\vspace{-1cm} 

\author{J. A. Aguilar-Saavedra}
\author{J. A. Casas}
\author{J. M. Moreno}
\affiliation{Instituto de F\'isica Te\'orica IFT-UAM/CSIC, c/Nicol\'as Cabrera 13--15, 28049 Madrid, Spain}

\begin{abstract}
For decades, it has been known that local hidden variable theories cannot be disproved by collider experiments involving decaying particles. However, if these theories satisfy a small set of mild assumptions, they become testable. In particular, they can be disproved using Bell-like inequalities for $\mu^+ \mu^-$ and $\tau^+ \tau^-$ pairs.
\end{abstract}

\maketitle

\section{Introduction}
High-energy collider experiments offer a new arena to probe quantum correlations, qualitatively different from usual low-energy experiments. At colliders, entangled states are produced dynamically in relativistic scattering and decay processes, usually involving unstable particles. Examples include top-quark pairs~\cite{Afik:2020onf,Severi:2021cnj,Afik:2022kwm,Aoude:2022imd,Aguilar-Saavedra:2022uye,Afik:2022dgh,Severi:2022qjy,Dong:2023xiw,Han:2023fci,Maltoni:2024tul,Maltoni:2024csn,Cheng:2024btk,Han:2024ugl,Aoude:2025ovu,Guo:2026yhz,Afik:2026pxv},  weak-boson pairs~\cite{Barr:2021zcp,Aguilar-Saavedra:2022wam,Ashby-Pickering:2022umy,Aguilar-Saavedra:2022mpg,Fabbrichesi:2023cev,Morales:2023gow,Aoude:2023hxv,Bernal:2023ruk,Fabbri:2023ncz,Aguilar-Saavedra:2024whi,Bernal:2024xhm,Ruzi:2024cbt,Grossi:2024jae,Wu:2024ovc,Bernal:2025zqq,DelGratta:2025qyp,Ding:2025mzj,Aguilar-Saavedra:2025byk,Goncalves:2025mvl,Goncalves:2025xer,Aguilar-Saavedra:2025njw,Pelliccioli:2026ltl,Aguilar-Saavedra:2026wuq}, lepton pairs~\cite{Altakach:2022ywa,Aguilar-Saavedra:2023lwb,Ehataht:2023zzt,Fabbrichesi:2024wcd,Han:2025ewp,Zhang:2025mmm}, $b$-quark pairs~\cite{Afik:2025grr}, as well as particles of different spin~\cite{Aguilar-Saavedra:2023hss,Aguilar-Saavedra:2024fig,Aguilar-Saavedra:2024hwd,Aguilar-Saavedra:2024vpd}

In current collider experiments it is not possible to perform direct spin measurements, let alone select measurement settings. This is so because the particles involved decay in a very short time, typically less than $10^{-12}$ s. Instead, the spin density matrices are reconstructed from angular distributions,  assuming the standard model (SM) for the matching between angular observables and spin observables. 
Such procedure allows one to test properties of the {\em quantum} state produced in the collision. 

This is a suitable approach to test quantum entanglement, which, by definition, is a {\em quantum} property that occurs when the state of a system cannot be defined without referring to the state of another system. In practice, 
one reconstructs from experiment —totally or partially— the (quantum) density matrix and applies the Peres-Horodecki criterion for entanglement~\cite{Peres:1996dw,Horodecki:1997vt} or an alternative one. Of course, such a procedure implies the assumption of quantum mechanics (QM). This is the strategy followed in recent ATLAS and CMS analyses that established entanglement in $t\bar t$ production~\cite{ATLAS:2023fsd,CMS:2024pts,CMS:2024zkc}.

Bell nonlocality, in contrast, is a statement about correlations between measurement outcomes in spatially separated subsystems and does not rely on the formalism of QM. Therefore, it is a notion distinct from entanglement. The point, of course, is that, unlike QM, any local hidden-variable theory (LHVT) must satisfy Bell inequalities, such as the Clauser–Horne–Shimony–Holt (CHSH) inequality~\cite{Clauser:1969ny}. A genuine test of Bell nonlocality requires performing independent measurements on each subsystem with randomly chosen settings, ensuring causal separation. This makes it impossible to perform loophole-free nonlocality tests in current collider experiments, as was early realised~\cite{Abel:1992kz}.

Certainly, for {\it quantum} systems in pure states, entanglement implies Bell nonlocality (and the other way around), but for mixed states it does not. E.g. for some Werner states~\cite{Werner:1989zz}, which are entangled, the results of all projective measurements can be described by a LHVT (and thus respect any Bell inequality), stressing the different nature of entanglement and nonlocality.

Even though fully genuine Bell tests at colliders are not feasible, it is possible and relevant to demonstrate that the {\em quantum} state produced in a collider experiment is nonlocal~\cite{Afik:2025ejh}. This entails reconstructing its spin density matrix with enough precision to assert that the joint state contains correlations that  would violate some Bell inequality. Furthermore, even when assuming QM for the reconstruction of spin states we may eventually disprove it, provided the correlations found are incompatible with the predictions of QM~\cite{Aguilar-Saavedra:2025cej}.

On the other hand, as it was shown in Ref.~\cite{Abel:1992kz},
the results of any experiment where the only measured observables are the momenta of the final particles ---as is the case in all current high-energy experiments--- can be reproduced by an explicit LHVT\footnote{More precisely, the LHVT is a version of the Kasday construction \cite{Kasday1971,Kasday1972} where the hidden variables are simply the momenta of the outgoing particles.}. Consequently, collider experiments cannot probe Bell inequalities without additional assumptions.

Between these two general facts about Bell nonlocality lies an interesting territory worth  exploring. Namely, it is possible to investigate under which assumptions, weaker than QM, it is possible to certify the violation of Bell inequalities in a collider. This is the goal of the present work.

\section{Spin correlations from momentum correlations.}
\label{sec: assumptions}

Since in a collider experiment one only measures the momenta of the outgoing particles, the only way to obtain information about spin correlations, and thus about the possible violation of Bell locality, is to relate them to momentum correlations of the decay products. This is actually quite straightforward if one assumes the validity of the SM and thus of QM. In the context of an LHVT this can be done using a small number of mild assumptions. This approach was first explored in Ref.~\cite{Bechtle:2025ugc}. 

For the sake of concreteness we will consider the case of two spin-$1/2$ particles, say $A$ and $B$, with respective decay products $a$ and $b$ (among others). 
The set of assumptions to establish the connection between the momentum correlation of the daughter particles and the spin correlation of the mother ones in an LHVT is the following:

\begin{enumerate}
    \item[i.] Poincaré invariance.

    \item[ii.] The decays of $A$ and $B$ are independent of each other.

    \item[iii.] The spin of each particle is an element of reality in the Einstein-Podolsky-Rosen~\cite{Einstein:1935rr} sense, i.e.  it is a vector with a definite orientation. The spins therefore play the role of hidden variables.

    \item[iv.] If the particle $A$ has spin $\hat{s}_A$, the probability distribution (in its rest frame) of the momenta of the daughter particles is always the same and  
    depends only on $\vec{s}_A$. The same holds for particle $B$. 
\end{enumerate}

Assumptions (i) and (ii) were stated in the same way in Ref.~\cite{Bechtle:2025ugc}, while assumption (iv) ---as well as (iii)--- was also implicitly used. 
Both (i) and (ii) are robust. In particular, the latter is a necessary consequence of a {\em local} HVT when the two decays are spacelike separated. Assumptions (iii) and (iv) are weaker, although quite plausible. Assumption (iv) implies that additional hidden variables different from the spins do not play any relevant role in the decay distributions. Specifically, for any two ensembles of particles $A$ having the same spin $\hat s_A$,
the decay distributions are the same.
As we will see, this is well supported by experimental data, at least for certain particles, such as muons or taus. A weaker version of assumption (iv) is discussed in appendix~\ref{sec:a}.

\begin{figure}[ht!]
   \begin{center}
  \fbox{\includegraphics[scale=0.5]{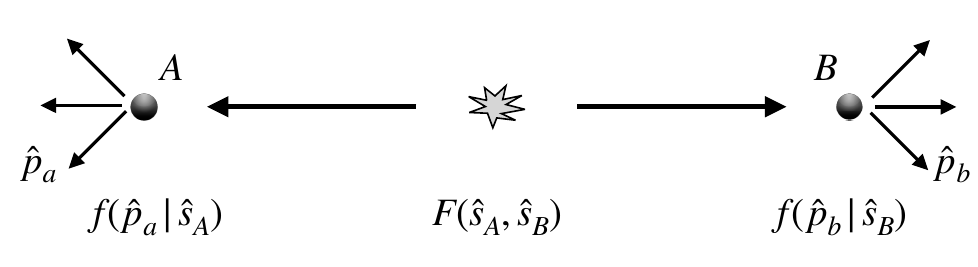}}
     \end{center}
	\caption{The two particles, $A$ and $B$, are produced with a certain probability distribution of the respective spins, $F(\hat s_A,\hat s_B)$. Upon decay, the daughter particles ($a$ and $b$) have momenta $\hat p_a$ and $\hat p_b$, with a probability distributions that depend only on the spin of the mother particle.
   }
    \label{fig:1}
\end{figure}

We now work out the relation between spin and momentum correlations. We define the momentum correlation of the daughter particles as
\begin{equation}
P_{ij} = \langle  p_{ai} p_{bj} \rangle = \int d\Omega_a d\Omega_b p_{ai} p_{bj} f(\hat p_a,\hat p_b) \,,
\label{ec:Pij}
\end{equation}
with $f(\hat p_a,\hat p_b)$ the probability density function (p.d.f.) to produce momenta along the unitary vectors $\hat p_a$, $\hat p_b$, i.e. the normalised differential cross section
\begin{equation}
\frac{1}{\sigma}\frac{d\sigma}{d\Omega_a d\Omega_b} = f(\hat p_a,\hat p_b) \,,
\end{equation}
and $\Omega_{a,b}$ the angles describing the orientation of $\hat p_{a,b}$.\footnote{To avoid confusion with other quantities, we use the letters $f$, $F$ to denote probabilities throughout the paper. We also use the standard notation $f(\cdot | \cdot)$ for conditional probabilities.}
Using the definition of conditional probability,
this p.d.f. can be written as
\bea
f(\hat p_a,\hat p_b) & = & \int d\bar\Omega_A d\bar\Omega_B \,  f(\hat p_a,\hat p_b | \hat s_A,\hat s_B)
\notag \\
& & \ \times\ F(\hat s_A,\hat s_B) \,,
\label{ec:f}
\eea
where $F(\hat s_A,\hat s_B)$ denotes the probability distribution of the (normalised) spins $\hat s_A$, $\hat s_B$, and $\bar\Omega_{A,B}$ the angles describing the orientation of these vectors. 
Assumption (iv), together with (ii), implies the factorisation (see Fig.~\ref{fig:1})
\begin{equation}
f(\hat p_a,\hat p_b | \hat s_A,\hat s_B) = f(\hat p_a | \hat s_A) f(\hat p_b | \hat s_B)\,.
\end{equation}
Hence,
\begin{eqnarray}
P_{ij} & = & \int d\bar\Omega_A d\bar\Omega_B F(\hat s_A,\hat s_B)
\left[ \int d\Omega_a p_{ai} f(\hat p_a | \hat s_A) \right] \notag \\
& & \times \left[ \int d\Omega_b p_{bi} f(\hat p_b | \hat s_B) \right] \,.
\label{ec:Pij2}
\end{eqnarray}
Because of rotational invariance, the functions $f(\hat p | \hat s)$  can only depend on $\hat p \cdot \hat s$ (we drop all subindices for brevity). Therefore,
we can expand them in terms of Legendre polynomials as
\begin{equation}
f(\hat p | \hat s) = \frac{1}{4\pi} \sum_{l=0}^{\infty} b_l P_l(\hat p \cdot \hat s) \,,
\label{ec:legexp}
\end{equation}
with $b_0 = 1$ by normalisation. Notice that in QM one has a similar expression, truncated above $l=1$. Plugging (\ref{ec:legexp}) into (\ref{ec:Pij2}) and applying the Funk-Hecke formula \cite{muller1998spherical}
\begin{equation}
\int d\Omega g(\hat p \cdot \hat s) Y_l^m(\Omega) = 2\pi Y_l^m(\bar \Omega) \int_{-1}^{1} dt\, g(t) P_l(t)\,, 
\end{equation}
where $\Omega$ and $\bar\Omega$ describe the orientation of $\hat p$ and $\hat s$, we find
\begin{equation}
P_{ij} = \frac{\alpha_a \alpha_b}{9}  \int d\bar\Omega_A d\bar\Omega_B F(\hat s_A,\hat s_B) \hat s_{Ai}  \hat s_{Bj} = \frac{\alpha_a \alpha_b}{9} C_{ij} \,,
\label{PC}
\end{equation}
Here $\alpha=b_1$, with the subscripts $a$ and $b$ referring to the distributions $f(\hat p_a | \hat s_A)$ and $f(\hat p_b | \hat s_B)$, respectively.
$C_{ij}$ is the average value of $s_{Ai} s_{Bj}$. The constant $\alpha$ is usually called spin analysing power in QM.
This formula, first obtained in Ref.~\cite{Bechtle:2025ugc}, shows the connection between momentum and spin correlations. We stress that its derivation relies on the four assumptions in the above list. 

Bell tests at colliders proceed by extracting the spin correlations $C_{ij}$ from experimental data on $P_{ij}$. If such spin correlations violate a  CHSH inequality, then the experiment cannot be described by a LHVT satisfying assumptions (i)--(iv) above.

It is interesting to note that the LHV construction formulated in  Refs.~\cite{Kasday1971, Abel:1992kz, Abel:2025skj} to reproduce 
the results of a collider experiment does not satisfy assumption (iv), except if $\hat p_{a,b}$ are fully determined by $\hat s_{A,B}$, which requires 
$f(\hat p_a | \hat s_A)=\delta(\hat p_a \cdot \hat s_A\pm 1)$, $f(\hat p_b | \hat s_B)=\delta(\hat p_b \cdot \hat s_B\pm 1)$.

\section{Continuous and binary observables}

Equation~(\ref{PC}) is formally identical to the one arising from QM,
\bea
P_{ij}=\frac{\alpha_a \alpha_b}{9} C_{ij}=\frac{\alpha_a \alpha_b}{9} \langle S_{Ai} S_{Bj}\rangle\,,
\eea
where $S_{Ai}$ ($S_{Bj}$) is the observable associated to the $i$ ($j$) component of the —normalised— spin of particle $A$ ($B$). However, there is an important difference. In QM the spin components are binary observables that take values $\pm 1$. In contrast, in the expression (\ref{PC}) the spin components are continuous variables. This assumption is embedded in the previous derivation of the formula. On the other hand, the usual CHSH inequality 
\bea
E(A_1 B_1)+E(A_1 B_2)+E(A_2 B_1)-E(A_2 B_2)\leq 2\,,
\eea
where $A_{1,2}$ ($B_{1,2}$) are Alice's (Bob's) observables and $E(\cdot)$ denotes mean value, holds for binary observables (taking values $\pm 1$ for convenience).

In order to relate the LHVT quantity $C_{ij}$ to the mean value of the product of two binary observables one has to give a prescription of how the values of $s_{Ai},s_{Bj}$ determine the output of a binary measurement of the spin; e.g. $S_{Ai}=+1$($-1$) if 
$s_{Ai}>0$ ($s_{Ai}<0$). Such prescription could be itself part of the LHVT. The problem is that, even with a defined prescription, there is no sufficient information in the continuous correlation $\langle s_{Ai}s_{Bj}\rangle$ to extract the binary correlation $\langle S_{Ai}S_{Bj}\rangle$ in an unambiguous way. 

Fortunately, such step is not necessary, since continuous variables are also subject to CHSH-like relations. As it is shown in Appendix~\ref{sec:b}, the extension of the CHSH inequality to continuous spin components reads:
\begin{align}
& E(s_{u_1} s_{v_1})+E(s_{u_1} s_{v_2})+E(s_{u_2} s_{v_1})-E(s_{u_2} s_{v_2}) \leq  \sqrt {2}
\nonumber\\
& \times   \left[ 1 + \left[(\hat u_1\cdot\hat u_2)^2 + (\hat v_1\cdot\hat v_2)^2  -  (\hat u_1\cdot\hat u_2)^2  (\hat v_1\cdot\hat v_2)^2 \right]^\frac{1}{2} \right]^\frac{1}{2},
\label{continuous CHSH}
\end{align}
where the unit vectors $\hat u_{1,2}, \hat v_{1,2}$ denote the directions along which the spin of $A$ and $B$ is measured with a perfect detector. 

Interestingly, the upper bound in (\ref{continuous CHSH}) is always less than or equal to 2. Therefore, if the spin correlations derived from Eq.~(\ref{PC}) violate the usual CHSH bound, they necessarily violate the continuous-variable version (\ref{continuous CHSH}). Moreover, if the chosen spin observables are not aligned, i.e. $|\hat u_1 \cdot \hat u_2|, |\hat v_1 \cdot \hat v_2| \neq 1$ ---as is typical in Bell tests--- the bound becomes even stronger than the standard CHSH one. In particular, for $|\hat u_1 \cdot \hat u_2| = |\hat v_1 \cdot \hat v_2|=0$, which is the optimal configuration when the two-particle system is in a maximally entangled state,  the upper bound reduces to $\sqrt{2}$, significantly below 2.

As it was stressed in Ref.~\cite{Bechtle:2025ugc}, one of the main obstacles to test Bell nonlocality in colliders is to determine the value of the coefficients $\alpha_a, \alpha_b$ entering the relation (\ref{PC}) between momentum and spin correlations without assuming the SM framework. We address this difficulty below. However, it is interesting to note in passing that for the Bell test we do not actually need that these coefficients are equal to (or smaller than) the ones derived from the SM. A slightly larger value would also work. E.g. if $A$, $B$ are a top quark and anti-quark produced at the LHC, and $a$, $b$ the daughter leptons, then the SM prediction is $|\alpha|=1$ for both decays. However, when the system is close to the singlet state a larger value, $|\alpha|\simlt 2^{1/4}$ would also work for the Bell test.

\section{Determination of $\alpha$}

As mentioned above, the relation (\ref{PC}) by itself is not enough to establish Bell violation in a collider experiment. Namely, the positivity constraint in the expansion (\ref{ec:legexp}) only sets a bound $|\alpha| =|b_1|\leq 3$. Hence, since $P_{ij}$ respects any Bell inequality, one cannot derive a violation in $C_{ij}$. This introduces the question of how to obtain experimental information about the $\alpha$'s.

Let us consider the decay of any particle.
In the context of a LHVT  satisifying assumptions (i)--(iv) of section \ref{sec: assumptions}, the
momentum p.d.f. (i.e. the normalised differential cross section) is
\begin{equation}
\frac{1}{\sigma}\frac{d\sigma}{d\Omega} =
f(\hat p) = \int d\bar \Omega f(\hat p | \hat s) F(\hat s) \,.
\label{ec:f1}
\end{equation}
where we denote by $\Omega = (\theta,\phi)$, $\bar \Omega = (\bar \theta,\bar \phi)$ the spherical coordinates of $\hat p$, $\hat s$ in an arbitrary reference system $(x,y,z)$. 
We can expand $F(\hat s)$ in spherical harmonics\bea
F(\hat s)=\sum_{l=0}^\infty
\sum_{m=-l}^l 
d_{lm}Y_{lm}(\bar \Omega). 
\eea
Then using the expansion (\ref{ec:legexp}) and applying the Funk-Hecke formula we get
\begin{equation}
f(\hat p) = \sum_{l=0}^\infty
\sum_{m=-l}^l \frac{1}{2l+1} b_l d_{lm}Y_{lm}(\Omega).
\label{distrib p}
\end{equation}
In order to study the correlation between $\hat p$ and the $\hat z$ axis we can integrate over the azimuthal angle $\phi$, $f_1(\cos \theta) \equiv \int_0^{2\pi} d\phi f(\hat p)$. This p.d.f. corresponds to the singly-differential cross section. 
\begin{equation}
\frac{1}{\sigma}\frac{d\sigma}{d\!\cos \theta} =
f_1(\cos \theta) = \frac{1}{\sqrt \pi} \sum_{l=0}^\infty \frac{b_l d_{l0}}{\sqrt{2l+1}} P_l(\cos \theta) \,.
\label{ec:f1final}
\end{equation}
The relevant point here is that the probability distribution (\ref{ec:f1final}) —which is measurable— depends on the products $b_l d_{l0}$, preventing the experimental extraction of the $b_l$ coefficients of the expansion (\ref{ec:legexp}) without additional information about the coefficients $d_{l0}$. This applies in particular to $b_1=\alpha$, which is required in the relation (\ref{PC}) to obtain the spin correlations from the momentum correlations, and thus the possible Bell-violation. Such limitation was pointed out in Ref.~\cite{Bechtle:2025ugc}; however, their treatment was incomplete. Equation~(50) of Ref.~\cite{Bechtle:2025ugc} is only valid when $\hat s$ lies in the directions $\hat z$ or $-\hat z$. Instead, Eq. (\ref{ec:f1final}) is the correct, general expression for the angular distribution with respect to a fixed axis.

There are two simple situations where information can be extracted from data in our framework. One of them is when the spin direction is known, $\hat s = \hat s_0$. Then, $f(\hat p) = f(\hat p | \hat s_0)$, and
\begin{equation}
f_1(\hat p \cdot \hat s_0) = \frac{1}{2} \sum_{l=0}^\infty b_l P_l(\hat p \cdot \hat s_0) \, .
\label{f1simplified}
\end{equation}
Measuring the angular distribution with respect to the axis $\hat s_0$ provides the value of $b_1=\alpha$, as well as higher-order coefficients. 

Another example is when we have two particles for which we know that the spins are (anti-)aligned, $\hat s_B = \pm \hat s_A$. The average value of the angle between the momenta of the daughter particles, $\hat p_a \cdot \hat p_b$ can be 
written in analogy to Eq.~(\ref{ec:Pij2}). If for example the spins are anti-aligned, then $F(\hat s_A,\hat s_B) = F(\hat s_A) \delta(\bar \Omega_a + \bar \Omega_B)$.  The integral on $\hat \Omega_B$ can be performed, yielding
\begin{eqnarray}
\langle \hat p_a \cdot \hat p_b \rangle & = & \int d\bar \Omega_A F(\hat s_A) \int d\Omega_a d\Omega_b \, \hat p_a \cdot \hat p_b \, f(\hat p_a | \hat s_A) \notag \\
& & \times\  f(\hat p_b | \shortminus\! \hat s_A) \,.
\label{ec:paxpb1}
\end{eqnarray}
We can parameterise $\hat p_a$, $\hat p_b$ with respect to the direction $\hat s_A$ in full generality. Then,
\begin{eqnarray}
\hat p_a \cdot \hat p_b & = & \frac{4\pi}{3} \left[ Y_1^0(\Omega_a) Y_1^0(\Omega_b) 
- Y_1^1(\Omega_a) Y_1^{-1}(\Omega_b) \right. \notag \\
& & \left. - Y_1^{-1}(\Omega_a) Y_1^1(\Omega_b) \right] \,.
\label{ec:paxpb}
\end{eqnarray}
The integral over momenta in (\ref{ec:paxpb1}) is straightforward. Because of the orthogonality of spherical harmonics, only the $l=1$ terms in the expansion of $f(\hat p | \hat s)$, c.f. (\ref{ec:legexp}), contribute. Furthermore, only the first term in (\ref{ec:paxpb}) survives after integration on azimuthal angles, and the integral yields a constant factor $-\alpha_a \alpha_b / 9$. The integral over $\bar \Omega_A$ is unity by normalisation, therefore 
\begin{equation}
\langle \hat p_a \cdot \hat p_b \rangle = - \frac{\alpha_a \alpha_b}{9} \,,
\end{equation}
allowing the measurement of the product $\alpha_a \alpha_b$. 

\section{Measuring $\alpha$ for lepton decays}

Let us apply the previous considerations to practical cases. An experimental determination of $\alpha$ for the decay of leptons $\ell = \mu,\tau$ is in principle possible, based on the following assumptions concerning particle properties:
\begin{enumerate}
\item[v.] Neutrinos have left-handed helicity, and anti-neutrinos right-handed helicity.\footnote{This statement is independent of whether neutrinos are Dirac or Majorana particles; antineutrinos are defined here as those produced together with a negative charged lepton.}
\item[vi.] Pions (for $\ell=\mu$), or mesons $D_s^\pm$/$B^\pm$ (for $\ell=\tau$) have spin zero.  
\end{enumerate}
Assumption (v) is founded on the classical experiment~\cite{Goldhaber:1958nb} of electron capture $^{152}\text{Eu}^* + e^- \to ^{152}\text{Sm}^* + \nu_e$. This experiment, and subsequent ones, verified that $\nu_e$ are left-handed by measuring the circular polarisation of photons emitted in the decay of $^{152}\text{Sm}^*$ to the ground state. Other neutrino flavours are assumed to also have left-handed helicity because of neutrino oscillations and Poincar\'e invariance. In the absence of a direct measurement of the anti-neutrino helicity, analogous to~\cite{Goldhaber:1958nb}, we assume that anti-neutrinos are right-handed from CPT symmetry.

Assumption (vi) for pions is supported by the detailed balance applied to the reaction $\pi^+ + d \leftrightarrow p + p$~\cite{Durbin:1951np} and the absence of fine structure in pionic atoms~\cite{Carrigan:1968sll},
among others. On the other hand, evidence for scalar nature is weaker for $D_s^\pm$ and $B^\pm$ mesons, and is based on the observed isotropy in the decays of these mesons.

From the previous mild assumptions, decays $\pi^\pm \to \mu^\pm \pbar{\nu}$ provide an example where the spin direction for muons is essentially known, based on angular momentum conservation. Therefore, one can use Eq.~(\ref{f1simplified}) to experimentally determine  $\alpha$ for $\mu^\pm$ decays. Likewise, $\alpha$ for $\tau$ decays could be determined in $D_s^\pm \to \tau^\pm \pbar{\nu}$. The determination of $\alpha$ needs knowledge of the flight direction of the decaying charged leptons, which may be difficult for $\tau^\pm$ due to their shorter lifetime.

We remark that if we assume that other mesons, for example $K^\pm$, also have spin zero, then the fact that muons from either $K^\pm \to \mu^\pm \pbar{\nu}$ or $\pi^\pm \to \mu^\pm \pbar{\nu}$ are observed to decay with the same p.d.f., supports assumption (iv) of section \ref{sec: assumptions}, that is, that the decay p.d.f. only depends on the spin of the parent particle. This observation also excludes the model described at the end of that section, where $\hat p\parallel \hat s$.

We briefly comment on the possibility that the $A$ and $B$ particles have (anti-)aligned spins. Consider, for example, decays $X \to A B$ where $X$ has spin zero~\cite{Pei:2026wfu}. One could conclude $\hat s_A = - \hat s_B$ in the absence of orbital angular momentum, $\vec L = 0$, which occurs if $X$ is a parity-odd scalar (not a parity-even scalar). However, to ascertain that $\vec L = 0$ from parity conservation, one has to assume the framework of QM.

\section{Testing locality with lepton pairs}

Testing locality with $\mu^+ \mu^-$ pairs is experimentally challenging, and unfeasible with current LHC detectors. With a lifetime of $2.2 \times 10^{-6}$ s at rest, their decay length is much above 1 Km for collider energies. This requires either placing far detectors (with the subsequent luminosity loss) or cooling the muons in such a way that coherence is preserved. We do not address here these experimental challenges.

An additional complication arises from the presence of multiple neutrinos in the decay $\mu^\pm \to e^\pm \nu \bar \nu$. These neutrinos prevent direct reconstruction of the $\mu^\pm$ rest frames unless the $\mu^\pm$ momentum is previously measured ---which may originate decoherence. Nevertheless, the momentum direction is easy to determine with an adequate experimental setup, and in that case one can test CHSH correlations in directions orthogonal to the $\mu^\pm$ momenta~\cite{Aguilar-Saavedra:2023lwb}. The correlations $P_{ij}$ in (\ref{ec:Pij}) are defined in the rest frame of the decaying particle, in this case the muons. Still, for directions orthogonal to the muon momentum the correlations $P_{ij}$ take the same value in the $\mu^+ \mu^-$ centre-of-mass frame. 

In contrast, tau leptons decay almost promptly, with a much shorter lifetime of $2.9 \times 10^{-13}$ s, so they represent a good option for Bell tests.
For $\tau^+ \tau^-$ pairs, there are proposals to reconstruct the rest frame in several production processes at various colliders~\cite{Altakach:2022ywa,Ehataht:2023zzt,Fabbrichesi:2024wcd,Zhang:2025mmm}. In any case, testing CHSH inequalities in directions orthogonal to the $\tau$ momentum may suffer from less uncertainties associated to the reconstruction.

\section{Interpretation and discussion}

In this work we have put forward a set of four, quite general,  assumptions on an LHVT that allow to relate spin correlations $C_{ij}$ of a pair of spin-$1/2$ particles $A$, $B$, to momentum correlations $P_{ij}$ of their decay products $a$, $b$. In this framework spin is a classical variable, described by a three-dimensional vector. The relation between $C_{ij}$ and $P_{ij}$ crucially depends on arbitrary quantities $\alpha_a$, $\alpha_b$ usually denoted as spin-analysing power of the daughter particles $a$, $b$. 
Then, we have identified two additional assumptions about particle properties that allows to experimentally measure $\alpha$ for muons and $\tau$ leptons without resorting to QM hypotheses.

We have derived CHSH-like inequalities that, unlike the original ones, are suited for such continuous variables. These inequalities could eventually be experimentally tested with $\mu^+ \mu^-$ or $\tau^+ \tau^-$ pairs, although the former is experimentally much more demanding.

In case a violation of these inequalities is found in experiment, then LHVTs satisfying the above assumptions would be excluded.

\section*{Acknowledgements}

We thank J.R.M. de Nova for useful comments.
This work has been supported by the Spanish Research Agency (Agencia Estatal de Investigaci\'on) through projects PID2022-142545NB-C21, PID2022-142545NB-C22 and CEX2020-001007-S funded by MCIN/AEI/10.13039/501100011033.

\appendix

\section{Alternative assumption (iv)}
\label{sec:a}

We can reconsider assumption (iv), explicitly addressing the role of hidden variables $\{\lambda\}$, which, together with the spins $\hat s_A, \hat s_B$, determine the decays of the $A$ and $B$ particles. We divide the set $\{\lambda\}$ into two subsets, $\{\lambda_A\}$ and $\{\lambda_B\}$, corresponding to the hidden variables that are relevant for the decay of the $A$ and $B$ particles, respectively (the intersection of both sets may be non-empty).

\begin{figure}[H]
   \begin{center}
  \fbox{\includegraphics[scale=0.5]{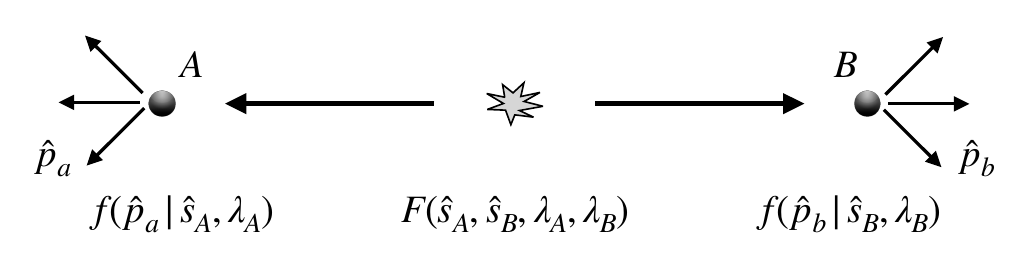}}
     \end{center}
	\caption{The same as Fig.~\ref{fig:1}, but now explicitly including the hidden variables $\lambda_A$ and $\lambda_B$.
   }
    \label{fig:2}
\end{figure}

The p.d.f. to produce momenta $\hat p_a$, $\hat p_b$ is
\begin{eqnarray}
f(\hat p_a,\hat p_b) & = & \int d\bar\Omega_A d\bar\Omega_B d \lambda_A d\lambda_B \,  f(\hat p_a,\hat p_b | \hat s_A,\hat s_B, \lambda_A, \lambda_B)
\notag \\
& & \ \times\ F(\hat s_A,\hat s_B, \lambda_A, \lambda_B) \,,
\label{ec:fv2}
\end{eqnarray}
where $F(\hat s_A,\hat s_B, \lambda_A, \lambda_B)$ is the probability distribution of the $\{\hat s_A,\hat s_B, \lambda_A, \lambda_B\}$ variables\footnote{In the case that a certain HV, say $\lambda^*$, belongs to both sets, $\{\lambda_A\}$ and $\{\lambda_B\}$, we can simply define two variables, $\lambda_A^*$ and $\lambda_B^*$, and include a factor $\delta(\lambda_A^*-\lambda_B^*)$ in the distribution function $F(\hat s_A,\hat s_B, \lambda_A, \lambda_B)$.}.
The requirement of locality ---assumption (ii)--- imposes the factorisation (see Fig.~\ref{fig:2})
\begin{equation}
f(\hat p_a,\hat p_b | \hat s_A,\hat s_B, \lambda_A, \lambda_B) = f(\hat p_a | \hat s_A, \lambda_A) f(\hat p_b | \hat s_B, \lambda_B)\,.
\label{ec:factfv2}
\end{equation}
We can now reformulate assumption (iv)  as follows:

\begin{enumerate}
\item[iv'.] 
For each particle pair, the hidden variables $\lambda_A$ are statistically independent of $\hat s_B$ and $\lambda_B$, and likewise $\lambda_B$ are statistically independent of $\hat s_A$ and $\lambda_A$.

\end{enumerate}

Its mathematical translation is the following. The probability distribution $F(\hat s_A,\hat s_B, \lambda_A, \lambda_B)$ can always be written as
\bea
F(\hat s_A, \hat s_B, \lambda_A, \lambda_B) = F(\lambda_A, \lambda_B|\hat s_A, \hat s_B) F(\hat s_A, \hat s_B)\,.
\label{prob cond}
\eea
Likewise,
\be
F(\lambda_A, \lambda_B|\hat s_A, \hat s_B)=F(\lambda_A|\hat s_A, \hat s_B)F(\lambda_B|\hat s_A, \hat s_B, \lambda_A)\,.
\ee
Then, assumption (iv') implies that the conditional probability factorises as
\bea
F(\lambda_A, \lambda_B|\hat s_A, \hat s_B) = F(\lambda_A | \hat s_A) F(\lambda_B | \hat s_B) \,,
\label{ec:factFv2}
\eea
where $F(\lambda_A | \hat s_A), F(\lambda_B | \hat s_B)$ are universal functions for all the $A$ and $B$ particles, respectively. As a consequence, after marginalising over the hidden variables $\lambda_{A,B}$, the probability distribution of $\hat p_a$ ($\hat p_b$) depends only on the spin of the parent particle, $\hat s_A$ ($\hat s_B$),
\bea
f(\hat p_a | \hat s_A)&=&\int d\lambda_A d\lambda_B f(\hat p_a | \hat s_A,\lambda_A) F(\lambda_A, \lambda_B|\hat s_A, \hat s_B)\nonumber\\
&=& \int d\lambda_A f(\hat p_a | \hat s_A,\lambda_A) F (\lambda_A | \hat s_A)
\eea
and similarly for $f(\hat p_b | \hat s_B)$.  
Now,
from Eqs.~(\ref{ec:fv2},\ref{ec:factfv2},\ref{ec:factFv2}),
\begin{eqnarray}
f(\hat p_a,\hat p_b) & = & \int d\bar\Omega_A d\bar\Omega_B F(\hat s_A,\hat s_B) \notag \\
& & \times \left[ \int d\lambda_A f(\hat p_a | \hat s_A,\lambda_A) F (\lambda_A | \hat s_A) \right] \notag \\
& & \times \left[ \int d\lambda_B f(\hat p_b | \hat s_B,\lambda_B) F (\lambda_B | \hat s_B) \right] \,.
\label{ec:fv2final}
\end{eqnarray}
The last two integrals are respectively $f(\hat p_a | \hat s_A)$ and $f(\hat p_b | \hat s_B)$.  Inserting (\ref{ec:fv2final}) into the definition of $P_{ij}$, Eq.~(\ref{ec:Pij}), we recover Eq.~(\ref{ec:Pij2}), from which the relation (\ref{PC}) between momentum correlations and spin correlations follows.

\vspace{0.2cm}

Finally, we show that without assumption (iv) or (iv'), omitted in Ref.~\cite{Bechtle:2025ugc}, one cannot guarantee that Eq.~(\ref{ec:Pij2}) ---and thus  Eq.~(\ref{PC})--- holds.

First of all, without the factorisation (\ref{ec:factFv2}), the momentum distribution of either daughter particle will, in general, depend on the spins of both $A$ and $B$,
\begin{equation}
f(\hat p_a | \hat s_A,\hat s_B) = \int d\lambda_A d\lambda_B f(\hat p_a | \hat s_A,\lambda_A) F(\lambda_A,\lambda_B | \hat s_A \hat s_B),
\label{pss}
\end{equation}
and similarly for $f(\hat p_b | \hat s_A,\hat s_B)$.
Exceptions may arise when, without relying on the factorisation (\ref{ec:factFv2}), the 
marginalised p.d.f.s
\begin{align}
&F(\lambda_A | \hat s_A , \hat s_B) = \int d\lambda_B F(\lambda_A,\lambda_B | \hat s_A \hat s_B) \,, \notag \\
& F(\lambda_B | \hat s_A , \hat s_B)=\int d\lambda_A F(\lambda_A,\lambda_B | \hat s_A \hat s_B) 
\label{marginalized}
\end{align}
do not depend on $\hat s_B$ and $\hat s_A$, respectively. In such case, 
$f(\hat p_a | \hat s_A)$ and $f(\hat p_b |\hat s_B)$ only depend on the spin of the parent particle. But even in this situation, it is not guaranteed that Eq.~(\ref{ec:fv2final}), and thus 
Eqs.~(\ref{ec:Pij2}, \ref{PC}), are recovered.

Let us illustrate this point with an example where $A$ and $B$ are the same particle species, and take $F(\lambda_A,\lambda_B | \hat s_A \hat s_B)$ to have the form
\begin{equation}
F(\lambda_A,\lambda_B | \hat s_A,\hat s_B) = G(\lambda_A,\hat s_A) \delta(\lambda_B - R(\hat s_A,\hat s_B) \lambda_A)\,,
\label{ec:Frara}
\end{equation}
where $G$ is a function that depends only on $\lambda_A$ and $\hat s_A$, and $R(\hat s_A,\hat s_B)$ is the rotation in the $(\hat s_A, \hat s_B)$ plane that maps $\hat s_A$ into $\hat s_B$, which is uniquely defined.  The action of rotations such as $R(\hat s_A,\hat s_B)$ onto the hidden variables $\lambda_A$ or $\lambda_B$ depends on their tensorial nature, which we leave unspecified (we only assume that the hidden variables transform covariantly under rotations). It follows that $F$ in (\ref{ec:Frara}) is invariant under arbitrary rotations $O \in \text{SO}(3)$ (as it should) provided $G(O \lambda_A,O \hat s_A) = G(\lambda_A,\hat s_A)$.

It is straightforward to verify 
that $F(\lambda_A | \hat s_A , \hat s_B) = G(\lambda_A,\hat s_A)$, $F(\lambda_B | \hat s_A , \hat s_B) = G(\lambda_B,\hat s_B)$. Hence, from (\ref{pss}) the momentum distributions $f(\hat p_a | \hat s_A)$, $f(\hat p_b | \hat s_B)$ only depend on the spin of the parent particle after hidden variables $\lambda_A$, $\lambda_B$ are marginalised. However, the momentum correlation is
\begin{eqnarray}
f(\hat p_a,\hat p_b) & = & \int d\bar\Omega_A d\bar\Omega_B F(\hat s_A,\hat s_B) \notag \\
\hspace{-1cm}
&& \times \left[ \int d\lambda_A G (\lambda_A ,\hat s_A)  f(\hat p_a | \hat s_A,\lambda_A)  \right.
\notag \\
&& \left.\phantom{\int} \times f(\hat p_b | \hat s_B,R(\hat s_A,\hat s_B)  \lambda_A) \right] \,,
\label{ec:fabrara}
\end{eqnarray}
and Eqs.~(\ref{ec:Pij2}, \ref{PC}) are not recovered.

\vspace{0.3cm}

\section{CHSH-like bound for continuous vector variables}
\label{sec:b}

We derive here Eq.~(\ref{continuous CHSH}) of the main text. From
\begin{equation}
 E(s_{\hat u} s_{\hat v})
 =
\int d\bar \Omega_A d\bar \Omega_B 
(\hat s_A \cdot\hat u) (\hat s_B \cdot \hat v)
F(\hat s _A, \hat  s_B) 
\end{equation}
we get
\begin{align}
 & E(s_{\hat u_1} s_{\hat v_1})+E(s_{\hat u_1} s_{\hat v_2})+E(s_{\hat u_2} s_{\hat v_1})-E(s_{\hat u_2} s_{\hat v_2})
 =  \nonumber\\[1mm] 
 & \int d\bar \Omega_A d\bar \Omega_B \hat s_A^T 
 {\cal M}  \hat s_B \,
F(\hat s _A, \hat  s_B) 
\end{align}
with ${\cal M} $ 
the $3 \times 3$ matrix
\begin{equation}
{\cal M} = 
 \hat u_1 \otimes
 (\hat v_1 +\hat v_2) + 
\hat u_2 \otimes (\hat v_1 -\hat v_2)\,.
\end{equation}
The ${\cal M}^T {\cal M}$ eigenvalues are given by
\begin{equation}
\left\{0,\  2 \pm 2\sqrt{(\hat u_1 \cdot\hat u_2)^2 + (\hat v_1\cdot\hat v_2)^2 -  
(\hat u_1 \cdot \hat u_2)^2  (\hat v_1 \cdot \hat v_2)^2} \right\} \notag\,.
\end{equation}
Since the probability density $ F(\hat s_A, \hat  s_B) $ is normalised, the largest eigenvalue provides the  bound  in Eq.~(\ref{continuous CHSH}).

\bibliographystyle{JHEP}
\bibliography{refs.bib}

\end{document}